Sequence Capture Versus Restriction Site Associated DNA Sequencing for Phylogeography

Michael G. Harvey[1,2], Brian Tilston Smith[2,3], Travis C. Glenn[4], Brant C. Faircloth[5], and Robb T. Brumfield[1,2]

[1]*Museum of Natural Science, Louisiana State University, Baton Rouge, LA 70803, USA*

[2]*Department of Biological Sciences, Louisiana State University, Baton Rouge, LA 70803, USA*

[3]Current address: *Department of Ornithology, American Museum of Natural History, Central Park West at 79th Street, New York, NY 10024, USA*

[4]*Department of Environmental Health Science, University of Georgia, Athens, GA 30602, USA*

[5]*Department of Ecology and Evolutionary Biology, University of California, Los Angeles, CA 90095, USA*

Corresponding author:

Michael G. Harvey

225-578-2855

mharve9@lsu.edu

ABSTRACT

Genomic datasets generated with massively parallel sequencing methods have the potential to propel systematics in new and exciting directions, but selecting appropriate markers and methods is not straightforward. We applied two approaches with particular promise for systematics, restriction site associated DNA sequencing (RAD-Seq) and sequence capture (Seq-cap) of ultraconserved elements (UCEs), to the same set of samples from a non-model, Neotropical bird. We found that both RAD-Seq and Seq-cap produced genomic datasets containing thousands of loci and SNPs and that the inferred population assignments and species trees were concordant between datasets. However, model-based estimates of demographic parameters differed between datasets, particularly when we estimated the parameters using a method based on allele frequency spectra. The differences we observed may result from differences in assembly, alignment, and filtering of sequence data between methods, and our findings suggest that caution is warranted when using allele frequencies to estimate parameters from low-coverage sequencing data. We further explored the differences between methods using simulated Seq-cap- and RAD-Seq-like datasets. Analyses of simulated data suggest that increasing the number of loci from 500 to 5000 increased phylogenetic concordance factors and the accuracy and precision of demographic parameter estimates, but increasing the number of loci past 5000 resulted in minimal gains. Increasing locus length from 64 bp to 500 bp improved phylogenetic concordance factors and minimal gains were observed with loci longer than 500 bp, but locus length did not influence the accuracy and precision of demographic parameter estimates. We discuss our results relative to the diversity of data collection methods available, and we provide advice for harnessing next-generation sequencing for systematics research.

New sequencing technologies promise to provide increasingly detailed estimates of species and population histories by resolving rapid radiations (Wagner et al. 2013), improving demographic parameter estimates (Jakobsson et al. 2008), and identifying regions of the genome under selection (Wang et al. 2009). However, it is still unclear which markers and library preparation methods are most useful for different research topics in systematics. Although continuing reductions in sequencing costs and advances in computational methods may eventually allow the widespread use of whole-genome sequencing for large comparative studies (Ellegren 2013), funding and computational constraints limit most researchers to sampling a subset of the genome (i.e., genome reduction) and multiplexing many subsampled individuals within single sequencing runs. Current strategies of genome reduction fall into four major categories: pooled PCR amplicon sequencing (Binladen et al. 2007), transcriptome or RNA sequencing (Morin et al. 2008), restriction site associated DNA sequencing (Baird et al. 2008), and sequence capture (Gnirke et al. 2009). Sequence capture and restriction site associated DNA sequencing hold particular promise for systematics because they 1) can use DNA extracts from tissues of low to moderate quality; 2) can be used with museum and genetic resource samples that do not contain high-quality RNA; 3) collect reasonably large amounts of data (i.e., thousands of loci) from many individuals (i.e., $\geq$100's) with modest effort; and 4) require relatively little specialized equipment.

Restriction site associated DNA sequencing (RAD-Seq) approaches involve digesting genomic DNA with one or more restriction enzymes, adding platform-specific adapters, and selecting fragments for sequencing that fall within a particular size distribution, effectively reducing the genome by sampling those regions near cut sites or where cut sites occur within a certain distance of one another (Baird et al. 2008). Variations on this general method differ primarily in the number of enzymes used (one or two), the types of enzymes used and the frequency of the targeted cut sites, whether random shearing is used on one end, and the approaches used for size selection (Davey et al. 2011; Stolle and Moritz 2013). In most restriction-digest-based methods, all fragments from a given locus have at least one static end (the cut site), meaning that sequence reads are not randomly distributed around a given cut site, which restricts the assembly of longer sequences from RAD-seq reads (Fig. 1). As a result, the general focus of RAD-Seq approaches is to generate single nucleotide polymorphism (SNP) data rather than longer sequences. SNPs from RAD-Seq have been widely used for genome-wide association studies, genome scans for selection, and other population genetic studies (reviewed in Narum et al. 2013). Restriction enzyme-based methods have been used to study phylogeographic and shallow phylogenetic questions (e.g., Eaton and Ree 2013; Wagner et al. 2013). RAD-seq methods are less useful for questions at deeper timescales, due to mutations in restriction sites among distantly related taxa and larger numbers of paralogous loci (Rubin et al. 2012).

Sequence capture (Seq-cap) approaches involve preparing DNA libraries from randomly fragmented DNA templates and hybridizing these libraries to synthetic oligonucleotide probes (60-120 mer) having sequence complementary to hundreds or thousands of genomic regions of interest. Large numbers of probes can be synthesized using microarray printing approaches and the resulting probes may then be used directly on the slides (Albert et al. 2007) or cleaved off the solid substrate and converted to biotinylated RNA using *in-vitro* transcription (Gnirke et al. 2009). Biotinylated RNA probes are currently more popular because RNA:DNA hybridization efficiency is higher in solution, fewer pieces of specialized equipment are necessary during the hybridization reaction (Gnirke et al. 2009), and RNA-probe based kits are commercially available from multiple vendors. Streptavidin-coated paramagnetic beads are used to pull down the biotin probes and hybridized (target) DNA library fragments, unwanted (non-target) portions of the DNA library are washed away, and targeted fragments are then released from the beads for sequencing (Gnirke et al. 2009; Fisher et al. 2011). Because enriched fragments are distributed randomly across targeted loci, the resulting sequence reads can be used for assembly of sequences longer than those assembled from RAD-seq data (Fig. 1).

Methods related to Seq-cap, involving the use of one to many oligonucleotides for targeted enrichment, have been in use for decades (e.g., Chong et al. 1993; Karagyozov et al. 1993; Kandpal et al. 1994). Recently, Seq-cap has been widely used for re-sequencing many loci in model systems for which genomes are available (e.g., Choi et al. 2009). In the absence of existing genomic resources, researchers can capture a large number of genomic regions using probes designed to target conserved sequences shared among divergent taxa (Faircloth et al. 2012; Lemmon et al. 2012; Hedtke et al. 2013; Li et al. 2013). For example, Faircloth et al. (2012) used sequence capture to target thousands of ultraconserved elements (UCEs), which are highly conserved, short, putatively non-coding sequences found throughout many animal genomes (Bejerano et al. 2004; Siepel et al. 2005). Phylogenetic studies using UCE loci have clarified a number of contentious relationships among mammals (McCormack et al. 2011), birds (McCormack et al. 2013), ray-finned fishes (Faircloth et al. 2013), and reptiles (Crawford et al. 2012). UCE loci contain enough variation in the regions flanking their conserved cores to resolve relationships at shallow, phylogeographic timescales (Smith et al. 2014).

Although both Seq-cap and RAD-Seq can be used for phylogeographic studies, the best method to use for a particular project likely varies depending on the taxa under study and the time-depth(s) of interest (Rubin et al. 2012). Potentially important factors affecting this decision also include differences between methods in cost (for equipment, supplies, computation, or labor); the amount of usable data generated; the availability of analytical tools compatible with resulting data; the accuracy of results from particular analyses using the resulting datasets; and the ease of comparing results across datasets, studies, taxa, and laboratories. Although either method may be used for phylogeographic studies, we are not aware of existing studies that directly compare Seq-cap and RAD-Seq data collected from the same samples or studies that summarize information about both methods to guide researchers interested in using one or the other approach for systematics research.

In this study, we examine phylogeographic datasets collected from the same set of Plain Xenops (*Xenops minutus*) samples using Seq-cap of UCEs and a RAD-Seq method called genotyping by sequencing (GBS; Elshire et al. 2011). *Xenops minutus* is a widespread bird inhabiting humid forests in Central America and northern South America (Parker III et al. 1996) that exhibits deep phylogeographic breaks across major barriers including the Andes Mountains (Smith et al. in review). We used both UCE Seq-cap and GBS RAD-Seq approaches to collect sequence read data, assemble loci, and call SNPs across all samples. We also generated several simulated Seq-cap and RAD-Seq datasets either of varying length or containing different numbers of loci. We use these empirical and simulated datasets to illustrate the utility of data collected using Seq-cap and RAD-Seq for phylogeographic analysis, determine the degree of concordance between estimates derived from Seq-cap and from RAD-Seq approaches, evaluate the accuracy of inferences based on simulated Seq-cap and RAD-Seq datasets having different attributes, and inform a general discussion of the application of Seq-cap and RAD-Seq to research in systematics.

METHODS

*Sampling*

We sampled four individuals of *X. minutus* from both west of the Andes Mountains (trans-Andes) and east of the Andes Mountains (cis-Andes; Fig. 1, Table S1). We extracted genomic DNA from tissue samples associated with voucher specimens using DNeasy tissue kits (Qiagen, Valencia, CA, USA) following the manufacturer's protocol.

*Genotyping by Sequencing (GBS) RAD-Seq*

We contracted the Cornell Institute of Genomic Diversity (IGD) to collect GBS RAD-Seq data using a standard workflow (Elshire et al. 2011). Briefly, the IGD digested DNA using PstI (CTGCAG) and ligated a sample-specific indexed adapter (eight total) and common adapter to resulting fragments. The IGD pooled and cleaned ligated samples using a QIAquick PCR purification kit (Qiagen), amplified the pool using an 18-cycle PCR, purified the PCR product using QIAquick columns, and quantified the amplified libraries using a PicoGreen assay (Molecular Probes, Carlsbad, CA, USA). Based on the PicoGreen concentration, the IGD then combined the eight pooled samples with 88 unrelated samples and ran the combined pool using a 100-base pair, single-end Illumina HiSeq 2000 lane.

The IGD processed raw sequence reads using the UNEAK pipeline, an extension to TASSEL 3.0 (Bradbury et al. 2007). Briefly, UNEAK retains all reads with an index, a cut site, and no missing data in the first 64 bases after the barcode. The UNEAK platform clusters reads having 100% identity into "tags", aligns tags pairwise, and calls tag pairs differing by one bp as SNPs. UNEAK outputs alleles represented by a minimum of five reads (heterozygotes must have >5 reads for each allele) and having a frequency of at least 5%. After receiving SNP data from IGD, we collapsed remaining reverse complement tag-pairs and re-called genotypes using the method of Lynch (2009) as implemented in custom PERL scripts obtained from Tom White (White et al. 2013) and available from https://github.com/mgharvey/misc (last accessed December 20, 2013). We also removed any SNPs for which genotype calls were missing for more than one of the eight individuals or that were heterozygous in more than six individuals to remove potential paralogs. We used the resulting SNP genotypes for analyses of the RAD-Seq SNP dataset. To generate the RAD-Seq sequence dataset, we added the consensus tag sequence preceding and following both alleles for each individual.

*Sequence Capture of Ultraconserved Elements (UCE Seq-cap)*

Details of the laboratory and data processing protocols we used are provided in Smith et al. (2013). In brief, we prepared libraries for eight samples using KAPA library preparation kits (Kapa Biosystems). We pooled all eight *X. minutus* samples in equimolar ratios, and we conducted sequence capture using 2,560 probes representing 2,386 UCEs following the workflow described by Faircloth et al. (2012; updates at http://ultraconserved.org, last accessed December 20, 2013), with several modifications to accommodate libraries prepared with KAPA kits. Prior to sequencing, we qPCR-quantified the enriched pool, combined this library pool with 36 unrelated libraries at equimolar ratios, and sequenced the combined libraries using one lane of a 100-base paired-end Illumina HiSeq 2000 run (Cofactor Genomics).

We demultiplexed raw reads using Casava 1.8 (Illumina, Inc.), quality filtered reads using Illumiprocessor (Faircloth 2011-12), which incorporates SCYTHE (Buffalo 2011-12) and SICKLE (Joshi 2013) for trimming adapter contamination and low-quality portions of reads. We conducted *de novo* assembly across all samples using VelvetOptimiser (Gladman 2009) and VELVET (Zerbino and Birney 2008). We aligned consensus contigs to UCE probe sequences and discarded un-aligned contigs and contigs matching multiple UCE loci using match_contigs_to_probes.py from the PHYLUCE package (Faircloth et al. 2012; https://github.com/faircloth-lab/phyluce, last accessed December 20, 2013). We mapped the cleaned reads for each individual back to an index of consensus contigs using BWA (Li and Durbin 2009), called SNPs for each individual and output consensus sequences using SAMtools (Li et al. 2009), and hard-masked low-quality bases (<Q20) using seqtk (Li 2013). We used MAFFT (Katoh et al. 2005) to generate final alignments for the sequence capture sequence dataset. For the Seq-cap SNP dataset, we randomly extracted a single polymorphic site from each alignment and we excluded alignments without any polymorphisms.

*Computation*

We used custom Python (van Rossum and Drake 2001) scripts (available at

https://github.com/mgharvey/misc) to generate input files for all subsequent programs. We

conducted sequence assembly and most analyses using compute nodes in the LSU High

Performance Computing cluster. Compute nodes included 2.93 GHz Quad Core Nehalem Xeon

64-bt processors with 24GB 1333 MHz RAM or 96GB 1066MHz RAM. We estimated population genetic summary statistics from SNP datasets using $\partial a\partial i$ (Gutenkunst et al. 2009) and from sequences using COMPUTE (Thornton 2003).

*Population Assignment and Admixture Estimation*

We used STRUCTURE (Pritchard et al. 2000) with the UCE Seq-cap and GBS RAD-Seq

SNP datasets to infer the most likely number of populations, assign individuals to populations,

and estimate admixture between populations. We tested each value of K between 1 and 8; ran

analyses for a 10,000 iteration burn-in followed by 1 million iterations; checked for convergence

by inspecting the alpha, D, and likelihood values; and compared the log probability of the data

across analyses to determine the most likely number of populations for either dataset (Pritchard

et al. 2000).

*Species Tree Estimation*

We were interested in determining the phylogenetic relationships between all samples in our analyses, including samples that clustered together in STRUCTURE analyses – a result that suggested low divergence and the potential for gene flow. Although many species tree methods assume that no migration occurs between populations (Heled and Drummond 2010; Bryant et al. 2012), concordance analysis should recover the primary vertical phylogenetic signal even in the presence of gene flow (Larget et al. 2010). Therefore, we examined topological concordance across loci and estimated species trees from both the UCE Seq-cap and GBS RAD-Seq sequence datasets using Bayesian concordance analysis (Baum 2007) implemented in the program BUCKy v1.4.2 (Larget et al. 2010). BUCKy estimates both a primary concordance tree that is a greedy consensus of the relationships supported by a large proportion of genes and a population tree based on quartet concordance factors that include branch lengths in coalescent units.

We used MrBayes v3.2.2 (Ronquist et al. 2012) to generate posterior distributions of gene trees for input to BUCKy. For each locus in either dataset, we conducted two independent runs of four chains in MrBayes using the GTR+$\Gamma$ substitution model. Based on ESS values and a visual inspection trace plots in Tracer v.1.5 (Rambaut and Drummond 2007), we determined that the analyses generally reached stationarity by 100,000 iterations for each tree. We ran each chain for 1,100,000 iterations, sampled trees every 500 iterations, discarded the first 200 trees as burn-in, and retained the remaining 2000 trees for concordance analysis. In BUCKy, we conducted analyses using different priors on the number of unique gene tree topologies (alpha=0.1, 1, 100), and we selected the optimization option to reduce memory requirements. For each analysis, we conducted two runs with four chains of 500,000 generations each. BUCKy analyses on the full RAD-Seq dataset failed due to insufficient memory on our servers (96GB RAM). To reduce memory requirements, we selected one of the two alleles from each individual for analysis. Because the tips in the BUCKy population tree contain a single allele, external branch lengths cannot be assigned using concordance factors, so we set these to one for ease of visualization.

*Demographic Analyses*

Computational methods for model-based demographic inference from SNP data and sequence data differ. We used either an approach designed for large SNP data sets ($\partial a \partial i$; Gutenkunst et al. 2009) or an approach designed for large, sequence data sets (G-PhoCS; Gronau et al. 2011) to estimate demographic parameters including divergence time, effective population sizes ($N_e$) of the ancestral and daughter populations, and migration rates between daughter populations. In all cases, we estimated parameters against a model in which an ancestral population diverged into two daughter populations (Fig. 3a), a model structure that is consistent with the results we inferred from STRUCTURE analyses of both RAD-Seq and Seq-cap datasets.

To estimate demographic parameters from Seq-cap and RAD-Seq SNP data, we used an approximate approach based on allele frequency spectra implemented in $\partial a \partial i$ (Gutenkunst et al. 2009). We were concerned that allele frequency spectra would differ between the datasets due to the different bioinformatics pipelines we used, particularly because we observed that more alleles were represented by a single allele copy (one allele in one individual) in the Seq-cap SNP dataset (36.3%) than in the RAD-Seq dataset (22.4%). As a result, we evaluated this potential source of error by analyzing both full frequency spectra as well as frequency spectra in which we masked singleton alleles. We masked singleton alleles by ignoring the portions of the frequency spectrum containing alleles that were only present in one copy in either the trans- or cis-Andean populations. For both masked and unmasked analyses, we used the diffusion approximation in $\partial a \partial i$ to simulate frequency spectra under the demographic model described previously. Based on a test of alternative optimization methods (Table S2), we selected the l-bgfs-b optimization routine to optimize parameter values and likelihoods. We observed that the program appeared to finish under local optima on occasion, so we ran each analysis five times and selected the run having the highest optimized likelihood value. We generated 100 conventional bootstrap replicates by sampling SNPs with replacement and again conducting five runs for each replicate to select the replicate run having the highest optimized likelihood.

To estimate demographic parameters from sequence data, we used the coalescent model and a Bayesian approach implemented in G-PhoCS (Gronau et al. 2011). We used G-PhoCS because of its ability to handle large datasets and because it can integrate over all possible phases of unphased diploid sequence data, allowing us to use alignments of unphased sequences as input. G-PhoCS uses a gamma ($\alpha$,$\beta$) distribution to specify the prior for the population standardized mutation rate parameter or theta ($\theta = 4N_e\mu$ for a diploid locus, where $\mu$ is per nucleotide site per generation), the population divergence time parameter ($\tau = T\mu$; $T =$ species divergence time in millions of years), and the migration rate per generation parameter ($m_{sx} \times \theta_x/4 = M_{sx}$), which is the proportion of individuals in population $x$ that arrived by migration from population $s$ per generation. Priors for $\tau$ and $\theta$ are adjusted using the shape parameter ($\alpha$) and scale parameter ($\beta$), which have a mean $\alpha/\beta$ and variance $s^2 = \alpha/\beta^2$. For $\tau$ and $\theta$, we examined the prior values (1, 30) and (1, 300), which represent wide ranges of divergence times and effective population sizes. We set the migration rate prior distribution parameters $\alpha$ and $\beta$ to 1.0 and 10, respectively. We ran each analysis twice for at least 400,000 generations and sampled the posterior distribution every 500 generations. We assessed MCMC convergence and determined burn-in by examining ESS values and likelihood plots using Tracer v.1.5 (Rambaut and Drummond 2007).

Without fossil or geological calibrations, converting parameters estimated from either Seq-cap or RAD-Seq data is not straightforward. We are not aware of accepted substitution rate estimates for UCEs or loci associated with PstI restriction sites. We instead scaled the cross-Andes divergence time estimated across all datasets to a divergence time estimate inferred from mitochondrial DNA (Burney and Brumfield 2009), which is thought to evolve in a relatively clock-like fashion in birds (Weir and Schluter 2008). We used this calibration to convert estimates of θ, migration rate, and substitution rate assuming a generation time of one year. We corrected θ values from the ∂a∂i analysis of Seq-cap data to correct for the fact that we only included a single SNP per locus.

*Analysis of Simulated Datasets*

Seq-cap and RAD-Seq datasets potentially differ in many ways, but we were interested in how phylogeographic analyses would be impacted by two important variables: the length of sequence generated and the number of loci targeted. We generated several simulated data sets consistent with empirical data from Seq-cap and RAD-seq approaches to examine the influence of different sequence lengths and numbers of loci on species tree inference and demographic parameter estimation. To simulate these data, we used a set of Python scripts that depend on ms (Hudson 2002), seq-gen (Rambaut and Grass 1997), and BioPython (Cock et al. 2009). These scripts are available as a fast and extensible approach for simulating next-generation sequencing datasets (https://github.com/mgharvey/mps-sim, last accessed December 20, 2013). We used these programs to simulate RAD-Seq and Seq-cap datasets containing eight diploid individuals divided into two extant populations. We selected a θ value of 0.4, tau of 0.4, and a low level of bidirectional migration ($N_{em} = 0.1$) between the two extant populations, approximating the values we inferred from the empirical datasets. We also adjusted gene tree scaling to simulate levels of variation similar to the empirical datasets ($\theta$/site = 0.00625 for RAD-Seq data; $\theta$/site = 0.002 for Seq-cap data). We discarded invariant loci and loci with greater than one SNP from simulated RAD-Seq datasets to approximate the methods used to generate the empirical RAD-Seq data. We simulated Seq-cap datasets containing 50% invariant sites and a gamma rate heterogeneity of alpha=0.5 to approximate the distribution of variation in UCE loci. We did not attempt to model patterns of missing data or sequencing error.

To examine the influence of sequence length on inference, we simulated Seq-cap datasets containing 500 alignments of 500, 1000, 5,000, and 10,000 bp in length and a RAD-Seq dataset containing 500 alignments of 64 bp in length. For each dataset, we inferred species trees using BUCKy and estimated demographic parameters in G-PhoCS using the same methodology we described for the empirical data.

To examine the influence of locus number on inference, we simulated RAD-Seq-like datasets containing 500, 1000, 5,000, 10,000, and 50,000 loci. For each dataset, we simulated alignments of 64 bp and selected only alignments containing a single biallelic SNP until we obtained the desired number of loci. We estimated species trees using BUCKy and demographic parameters using G-PhoCS and $\partial$a$\partial$I following the same methodology described above.

RESULTS

After processing and filtering, GBS RAD-Seq produced genotypes from 4,250 SNPs from unique loci and UCE Seq-cap resulted in sequence alignments averaging 604.2 bp in length from 1,368 loci containing 7,658 total SNPs. Table 1 provides additional attributes of the SNP and sequence datasets generated from both methods. We present summary statistics separately for both populations west of the Andes (trans-Andes) and east of the Andes (cis-Andes; Table 2). Estimates from SNPs are single values calculated from the sample-wide frequency spectrum, whereas estimates from sequences summarize values across loci. Standard deviations of sequence-based summary statistics are large, reflecting the large variance in the amount of variation across loci. Watterson's θ and nucleotide diversity estimates were higher in the RAD-Seq dataset than in the Seq-cap dataset when examining either the sample-wide values estimated from SNPs or the mean value across loci estimated from sequence data (although standard deviations of estimates from sequence data were largely overlapping). The estimates of Tajima's D derived from UCE Seq-cap SNP data were negative on both sides of the Andes, whereas Tajima's D estimates from the GBS RAD-Seq data were both positive. Standard deviations of estimates of Tajima's D from both sequence datasets overlapped zero.

*Population Assignment and Admixture*

The log likelihood of the data from STRUCTURE analyses with different K-values plateaued at K=2 when we used either sequence capture SNP or RAD-Seq SNP datasets, suggesting that individuals we sampled are best clustered into two populations. In both analyses, STRUCTURE assigned trans-Andean and cis-Andean individuals to different clusters with 100% probability. We did not detect admixture in analyses of either UCE or GBS dataset, and we assigned samples to each of these two populations, separated by the Andes, for subsequent analyses.

*Species Tree Estimation*

Species tree topologies inferred by BUCKy were identical, and branch lengths and concordance factors (concordance in support for each node across gene trees) were similar across alpha values, so we focus on results from the run with an intermediate alpha value (alpha=1). Within the RAD-Seq (Fig. 3a,b) or Seq-cap (Fig. 3c,d) datasets, BUCKy concordance trees and population trees shared identical topologies. Both concordance (Fig. 3a,c) and population (Fig. 3b,d) trees were topologically similar between RAD-Seq and Seq-cap datasets, with one poorly supported conflict. In the Seq-cap tree, one of the cis-Andean individuals is sister to a clade containing all of the other cis-Andean individuals, whereas in the RAD-Seq tree this individual is sister to just one of the other cis-Andean individuals. Concordance factors for relationships within the cis-Andean clade were low (<0.5) in analyses of both RAD-Seq and Seq-cap datasets, suggesting poor support for relationships inferred among those individuals. All concordance and population trees closely resemble a mitochondrial gene tree generated as part of a separate study (Fig. 3e).

*Demographic Parameter Estimation*

Demographic parameters inferred from $\partial a \partial i$ analyses of SNPs from RAD-Seq and Seq-cap datasets were dramatically different (Table 3). Mean effective population size estimates were more than two orders of magnitude larger in the Seq-cap results relative to the RAD-Seq results. When we masked singleton alleles (alleles present in one copy in either the trans- or cis-Andean populations), effective population sizes were somewhat more concordant between RAD-Seq and Seq-cap datasets. Migration rate estimates were very low for both datasets in both the masked and unmasked analyses. Estimated substitution rates were higher for RAD-Seq than Seq-cap data in analyses of the unmasked datasets, but lower for RAD-Seq than Seq-cap in analyses of the masked datasets.

In contrast to the $\partial a \partial i$ analyses of SNP data, parameter estimates derived from RAD-Seq and Seq-cap analyses of sequence data using G-PhoCS were similar (Table 3). The G-PhoCS estimate of ancestral effective population size was higher for RAD-Seq data than Seq-cap data, the estimates of trans-Andean effective population size were the same between datasets, and the estimates of cis-Andean effective population size was lower for RAD-Seq data than Seq-cap data. Estimates of migration rate and substitution rate from RAD-Seq and Seq-cap data had overlapping confidence intervals.

Between model-based, inferential approaches, effective population size estimates were generally lower in the G-PhoCS analyses than the $\partial a \partial i$ analyses. Migration rates estimated from sequence data using G-PhoCS were much higher than migration rates estimated from SNP data using $\partial a \partial i$. Substitution rates estimated across datasets and analyses were similar, although the estimates from three of the four $\partial a \partial i$ analyses were higher.

*Analyses of Simulated Data*

Simulated datasets of 500, 1000, 5000, and 10,000 bp in length averaged 4.7 (SD=2.4), 9.9 (SD=4.0), 48.2 (SD=13.6), and 98.7 (SD=26.3) polymorphic sites per locus, respectively. We successfully analyzed datasets of each length and number of loci using BUCKy, and we analyzed datasets with varying numbers of loci using $\partial a \partial i$. G-PhoCS analyses of the largest number of loci (50,000) failed, but all other treatments of locus length and number of loci completed successfully.

We evaluated the accuracy of species tree inference across datasets by measuring the quartet concordance factor for the split between the two daughter populations in the primary concordance tree. Species tree accuracy increased as alignment length increased from 64 to 500 bp, but the increase in species tree accuracy plateaued for alignment lengths greater than 500 bp (Fig. 4a). Among treatments varying the number of loci, species tree accuracy increased when we increased the number of loci from 1000 to 5000 (Fig. 4b), but we did not observe additional effects of including more loci.

To evaluate parameter estimates, we used a coefficient of variance ((standard deviation/mean)$^2$) as an index of parameter estimate precision and we used a point estimate (ML estimate or posterior mean) divided by the expected value used for simulation as an index of accuracy. Parameter estimates from $\partial a \partial i$ generally increased in accuracy and became more precise with the addition of more loci (Fig 4c,d). Migration rate estimates, however, were generally less accurate and precise, and did not show obvious trends, with increasing locus length or number (Table S4). Parameter estimates from G-PhoCS also did not show obvious trends in accuracy or precision as the length or number of loci increased, at least within the range of treatments that we examined (Table S5). Migration rate estimates from G-PhoCS had particularly low accuracy and precision (Table S5).

DISCUSSION

We successfully applied two methods of collecting next-generation sequencing data, Seq-cap of UCEs and GBS RAD-Seq, to the phylogeography of a non-model bird species, building upon several studies that have previously demonstrated the utility of both approaches for phylogeography (McCormack et al. 2012; Carstens et al. 2013; Narum et al. 2013; Smith et al. 2013). We used sequence data and SNPs derived from both methods to estimate summary statistics, conduct population assignment, estimate species tree topologies, and estimate demographic parameters. In addition, we showed that analyses of both Seq-cap and RAD-Seq datasets clustered sampled individuals into two populations with identical individual assignments and no admixture, and we found that primary concordance and population trees estimated from both methods were similar.

Demographic inferences from the full coalescent analysis using G-PhoCS were similar between Seq-cap and RAD-Seq datasets, but results inferred from $\partial a \partial i$ using the allele frequency spectrum were different. The differences we observed among parameters estimated using $\partial a \partial i$ may be due to local differences between the types of genomic loci examined (RAD versus UCE) or processing differences among bioinformatic pipelines. For example, to reduce the effect of paralogs, we removed loci from the RAD-Seq data set if either allele had fewer than five reads or a frequency of less than 5%, but we did not implement a similar filter while processing Seq-cap data. As a result, filtering may have removed real variation, variation introduced by the inadvertent combination of paralogs, and variation introduced by sequencing errors, potentially biasing the marginal spectra of daughter populations in the RAD-Seq dataset and leading to discrepancies we observed among effective population size estimates. We were also concerned that singleton alleles were potentially biasing results, but reanalyzing the datasets with singletons masked only marginally improved concordance between analyses, suggesting that the influence of differences in assembly and filtering are more pervasive across the frequency spectrum. Assembly and filtering approaches similar to those that we used are common among RAD-Seq studies lacking reference genomes (e.g., Rubin et al. 2012; Wagner et al. 2013; White et al. 2013) and are useful for mitigating problems introduced by paralogs and sequencing errors. However, in low-coverage datasets, such as those we analyzed, discriminating between errors and rare alleles is challenging (Nielsen et al. 2012; Fumagalli et al. 2013). Methods that help correct biased frequency data are available (Nielsen et al. 2012), but these generally depend on an existing panel of high-confidence reference SNPs from the same populations for comparison. Because reference panels are generally unavailable for non-model species and because filtering approaches can significantly affect RAD-seq data, researchers should be careful when making inferences based on frequency spectra in these systems.

One benefit of the RAD-Seq approach is that laboratory methods can be adjusted to target different numbers of loci (Baird et al. 2008; Peterson et al. 2012) and clustering thresholds can be adjusted to generate alignments with different numbers of SNPs (Baxter et al. 2011). Similarly, Seq-cap approaches can include capture baits or probes to target more or fewer parts of the genome and shorter or longer regions of the genome (Bi et al. 2012; Faircloth et al. 2012; Hedtke et al. 2013). We used simulated data to evaluate the effect of locus number and locus length on inferences using the methods we applied to our empirical data. Increasing the number and length of loci generally improved species tree estimation, but we observed diminishing returns at lengths greater than 500 bp or when including more than 5000 loci. Similarly, increasing the number of loci improved estimation of demographic parameters using the allele frequency spectrum ($\partial a \partial i$) approach, but these returns diminished above 5000 loci. Our findings are supported by previous studies that have recommended increasing both locus number and length to improve species tree estimation (Camargo et al. 2012; Harris et al. In press) or increasing the number of loci samples to improve the estimation of θ (Felsenstein 2006; Carling and Brumfield 2007). Our findings also suggest that increasing the number of loci over 5000 or the locus length over 500 bp yields diminishing returns in terms of improved species tree and demographic parameter estimation. Additional simulation-based studies investigating wider ranges of input parameters that might be achieved with next-generation sequencing and different analytical approaches that can be applied to large datasets are warranted.

*Seq-cap vs. RAD-Seq: Practical Considerations*

Although the number and length of loci sampled is an important component of designing any given study, many other factors influence the genomic sampling strategy. Financial, computational, and time investment factors are also important aspects to consider when selecting approaches to collect data (Sboner et al. 2011). Although next-generation sequencing platforms have dramatically reduced the cost and time involved in sequencing (Wetterstrand 2013; Glenn 2011), funding and time may still be limiting in large comparative studies due to expensive library preparations and limitations on the number of samples that can be multiplexed on a single sequencing lane (Harris et al. 2010). We generated Seq-cap and RAD-Seq datasets for this study at a similar cost per sample (~$60 US for Seq-cap and ~$40 US for RAD-Seq). Both methods can be conducted largely using equipment that is standard in most molecular labs (Gnirke et al. 2009; Elshire et al. 2011), so the cost of equipment purchase is largely negligible. Seq-cap will generally be more expensive than RAD-Seq due to the costs associated with more involved library preparation and with target enrichment (synthesizing oligonucleotide probes, streptavidin beads, secondary amplification and purification). Seq-cap requires greater sequencing depth and thus higher sequencing cost than RAD-Seq per locus, but this is offset because it more efficiently targets informative, single-copy loci (vs. more multi-copy and/or invariant loci in RAD-Seq). Thus, both Seq-cap and RAD-Seq can be conducted on a hundred samples for a few thousand US dollars (exclusive of labor). Similarly, time investment is modest for both methods (Gnirke et al. 2009; Elshire et al. 2011), although again Seq-cap is slower due to the more intensive library preparations and additional hybridization and enrichment steps. For about one hundred samples, library preparation for RAD-Seq can be completed in about two days, whereas an equivalent number of Seq-cap libraries can be prepared in about five days. Commercial sequencing services that offer library preparation, enrichment, and sequencing as a service are available for both methods.

Computational investment is a particularly important practical consideration in study design for next-generation sequencing projects. The ability to generate genetic data has outstripped the availability of methods to process sequencing reads and analyze the large datasets produced (Delsuc et al. 2005; Horner et al. 2010). We selected methods that have been developed specifically for genomic datasets. Even so, we had to deal with long run times (days to weeks) and high memory requirements (sometimes >48GB RAM). Depending on the question being addressed, very large datasets may not be needed and additional data may unnecessarily complicate analyses (Davey et al. 2011). Conversely, extremely challenging evolutionary events may require large amounts of data to yield reliable results, and larger datasets also offer the ability to subsample loci informing a research question *post-hoc*.

*Seq-cap vs. RAD-Seq: Data Processing and Dataset Attributes*

In this study, we demonstrated that both Seq-cap and RAD-Seq can produce informative datasets for phylogeography, but both methods also suffer from several limitations. Many RAD-Seq analysis pipelines, for example, require that similarity thresholds are set to cluster orthologous reads while separating or removing similar reads from different loci (Elshire et al. 2011). However, this procedure effectively normalizes the amount of variation recovered across datasets. As a result, patterns of variation across species and datasets may be comparable, but metrics of absolute variation are not. In addition, substitutions in restriction site recognition sequences within species create null alleles and substitutions between species limits the extent to which orthologous loci might be recovered across species (Rubin et al. 2012). Differences in methylation among individuals as well as different strategies for size selection or other procedures across protocols or laboratories may further reduce reproducibility across studies.

Seq-cap approaches to enrich UCEs or other conserved regions benefit from the presence of a (usually) single-copy, conserved region at each locus which reduces paralogy and lowers the rate of allelic dropout. However, in most sequence capture experiments (a) many reads (often about 70%) are lost to "off-target" areas and (b) the coverage distribution is centered on the probes, which for UCEs is the middle of conserved targets rather than the more variable flanking regions containing informative SNPs (Fig. 1; Faircloth et al. 2012; Smith et al. 2013).

Seq-cap of UCEs may be preferable for generating datasets that are explicitly comparable across species or deeper taxonomic groups, for example comparative phylogeography or phylogenetics. RAD-Seq may be preferable when researchers wish to maximize the amount of variation recovered within a single species and do not need to draw comparisons to other datasets, particularly when the study species diverged recently or likely contains low genetic variation.

Seq-cap and RAD-Seq may also be appropriate for different types of research questions due to the differences between the data each generates. Based on this study, the SNPs and short sequences generated by RAD-Seq maximize the number of loci recovered, but these short sequences result in low per-locus information content. Short, less-informative loci may preclude reconstructing informative gene trees and complicate analyses using full coalescent methods (Heled and Drummond 2010; Bayzid and Warnow 2013). Conversely, Seq-cap results in the recovery of much longer sequences and resulting alignments, but because sequence length is a function of high coverage, very long (>1000 bp) loci may come at the expense of the total number of loci recovered. Variations on RAD-Seq may result in longer loci (Davey et al. 2011), whereas other Seq-cap probe sets might contain more loci (see http://ultraconserved.org). Our simulation results highlight the fact that different analyses may benefit differently from increasing the number of loci or the length of sequences.

*Seq-cap vs. RAD-Seq: Analyses and Accuracy*

We demonstrated that both Seq-cap and RAD-Seq data can be analyzed using a variety of phylogeographic methods, but one or the other may be better for certain applications. RAD-Seq generates data from systematically cut fragments with recognition sequences that are often semi-randomly dispersed across the genome (Elshire et al. 2011), although some flavors of RAD-Seq produce fragments with more random distributions than others (Stolle and Moritz 2013). Thus, RAD-Seq may be more appropriate for studies involving whole genome scans particularly where genome resequencing is not feasible, for example to locate areas of potential interest in the study of adaptation (e.g., Andrew and Rieseberg 2013; Parchman et al. 2013). Alternatively, Seq-cap has the ability to generate data from most regions of the genome, provided prior sequence information is available (Gnirke et al. 2009). New or existing probe arrays can easily be augmented with probes targeting historically used loci, barcoding loci, new loci found to be informative with respect to demographic history, or candidate loci potentially important for adaptation. Sequence capture probe sets are also effective for obtaining data from ancient DNA samples (Bi et al. 2012).

The accuracy of estimates derived from RAD-Seq or Seq-cap data is another consideration. The evolution of the loci targeted in this study, PstI restriction sites and UCEs, is still poorly understood. Mutational and coalescent variance across loci (Maddison 1997; Huang et al. 2010), sequencing error (Fumagalli et al. 2013), and the fit of the data to the models examined (Reid et al. In press) may influence accuracy in either type of data. UCEs may be under strong purifying selection (Bejerano et al. 2004; Katzman et al. 2007), which may confound results pending the development of more accurate evolutionary models. There is debate about whether targeting conserved genomic regions is advantageous for phylogenomic inference (Betancur R. et al. In press) or not (Salichos and Rokas 2013). Future work should focus on developing more accurate evolutionary models for the regions of the genome from which we can collect data, as well as evaluating the fit of real datasets to commonly used models of sequence evolution.

*Seq-cap vs. RAD-Seq: The Importance of Deep Orthology*

One potentially important difference between RAD-seq and Seq-cap of UCEs is that Seq-cap of UCEs permits the collection of the same set of loci across divergent groups (Faircloth et al. 2012). This degree of orthology among captured targets allows researchers to investigate questions at different timescales, spanning population genetics to deep phylogeny. This conservation also allows the direct comparison of variation within and among different species (Smith et al. 2014). It is unclear whether examining results from sufficient numbers of different loci in different species may bias comparisons between them (Kuhner et al. 1998; Beerli and Felsenstein 1999; Carling and Brumfield 2007), but the ability to examine the same loci in each species removes this potential source of error. Perhaps the greatest potential benefit of UCEs and similar markers is that generated datasets may be easily incorporated into larger comparative studies. Emergent patterns from comparative studies using the same loci within different species or at different timescales may provide novel insight into the history of molecular evolution across time and space (Faircloth et al. 2012).

*Conclusions*

Seq-cap of UCEs and RAD-Seq can provide large amounts of information for phylogeography. Our results suggest that data from both methods can be used in diverse phylogeographic analyses, and our resulting estimates are largely concordant between both approaches and when compared with mitochondrial data. RAD-Seq is best for generating SNP data from many loci, and RAD-Seq approaches may be more appropriate for studies involving whole genome scans. Seq-cap of UCEs is best for generating longer sequences containing linked

SNPs, and may be more appropriate for downstream comparative phylogeography and phylogenetic studies. Both methods should be useful for population genetics and demographic inference, although better methods for dealing with biases introduced by processing low-coverage sequencing data are required. Of the analyses examined here, species tree inference is improved both by increasing the number of loci and number of linked SNPs examined, whereas demographic inference is improved by sequencing more loci.

DATA AVAILABILITY

All empirical and simulated datasets are available from Dryad (doi:10.5061/dryad.80s71).

AUTHOR CONTRIBUTIONS

Conceived and designed the study: MGH, BTS, TCG, BCF, RTB. Conducted laboratory work: MGH, BTS, TCG. Analyzed data and conducted simulations: MGH. Contributed reagents/code: MGH, BTS, TCG, BCF, RTB. Wrote the paper: MGH, BTS, TCG, BCF, RTB.

ACKNOWLEDGEMENTS

D. Willard (Field Museum), M. B. Robbins (University of Kansas Natural History Museum), and D. L. Dittmann (Louisiana State University Museum of Natural Science) provided genetic samples. J. M. Brown and B. C. Carstens discussed experimental design. C. Locklear at Integrated DNA Technologies (IDT) provided adapters and sequencing. B. J. Nelson provided programming advice and S. A. Taylor and T. A. White provided scripts and assistance with RAD-Seq data processing. I. Gronau and R. N. Gutenkunst, developers of G-PhoCS and $\partial a \partial i$, provided advice on analyses and parameter conversion. B. Thakur assisted with use of the computing resources at the LSU High Performance Computing facilities. This work was funded in part by NSF grants DEB-0841729 and DEB-1210556 (a Doctoral Dissertation Improvement Grant for MGH's dissertation) to RTB and DEB-1242267 to BCF and TCG.

REFERENCES

Albert, T. J., M. N. Molla, D. M. Muzny, L. Nazareth, D. Wheeler, X. Song, T. A. Richmond, C. M. Middle, M. J. Rodesch, and C. J. Packard. 2007. Direct selection of human genomic loci by microarray hybridization. Nat. Methods 4:903-905.

Andrew, R. L., and L. H. Rieseberg. 2013. Divergence is focused on few genomic regions early in speciation: incipient speciation of sunflower ecotypes. Evolution 67:2468-2482.

Baird, N. A., P. D. Etter, T. S. Atwood, M. C. Currey, A. L. Shiver, Z. A. Lewis, E. U. Selker, W. A. Cresko, and E. A. Johnson. 2008. Rapid SNP discovery and genetic mapping using sequenced RAD markers. PLoS One 3:e3376.

Baum, D. A. 2007. Concordance trees, concordance factors, and the exploration of reticulate genealogy. Taxon 56:417-426.

Baxter, S. W., J. W. Davey, J. S. Johnston, A. M. Shelton, D. G. Heckel, C. D. Jiggins, and M. L. Blaxter. 2011. Linkage mapping and comparative genomics using next-generation RAD sequencing of a non-model organism. PLoS One 6:e19315.

Bayzid, M. S., and T. Warnow. 2013. Naive binning improves phylogenomic analyses. Bioinformatics 29:2277-2284.

Beerli, P., and J. Felsenstein. 1999. Maximum-likelihood estimation of migration rates and effective population numbers in two populations using a coalescent approach. Genetics 152:763-773.

Bejerano, G., M. Pheasant, I. Makunin, S. Stephen, W. J. Kent, J. S. Mattick, and D. Haussler. 2004. Ultraconserved elements in the human genome. Science 304:1321-1325.

Betancur-R., R., G. Naylor, and G. Orti. In press. Conserved genes, sampling error, and phylogenomic inference. Syst. Biol. doi:10.1093.

Bi, K., D. Vanderpool, S. Singhal, T. Linderoth, C. Moritz, and J. Good. 2012. Transcriptome-based exon capture enables highly cost-effective comparative genomic data collection at moderate evolutionary scales. BMC Genomics 13:403.

Binladen, J., M. T. P. Gilbert, J. P. Bollback, F. Panitz, C. Bendixen, R. Nielsen, and E. Willerslev. 2007. The use of coded PCR primers enables high-throughput sequencing of multiple homolog amplification products by 454 parallel sequencing. PLoS One 2:e197.

Bradbury, P. J., Z. Zhang, D. E. Kroon, T. M. Casstevens, Y. Ramdoss, and E. S. Buckler. 2007. TASSEL: software for association mapping of complex traits in diverse samples. Bioinformatics 23:2633-2635.

Bryant, D., R. Bouckaert, J. Felsenstein, N. A. Rosenberg, and A. RoyChoudhury. 2012. Inferring species trees directly from biallelic genetic markers: Bypassing gene trees in a full coalescent analysis. Mol. Biol. Evol. 29:1917-1932.

Buffalo, V. 2011-12. Scythe. Available from https://github.com/vsbuffalo/scythe (last accessed December 20, 2013).

Burney, C. W., and R. T. Brumfield. 2009. Ecology predicts levels of genetic differentiation in Neotropical birds. Am. Nat. 174:358-368.

Camargo, A., L. J. Avila, M. Morando, and J. W. Sites. 2012. Accuracy and precision of species trees: effects of locus, individual, and base pair sampling on inference of species trees in lizards of the Liolaemus darwinii group (Squamata, Liolaemidae). Syst. Biol. 61:272-288.

Carling, M. D., and R. T. Brumfield. 2007. Gene sampling strategies for multi-locus population estimates of genetic diversity ($\theta$). PLoS One 2:e160.

Carstens, B. C., R. S. Brennan, V. Chua, C. V. Duffie, M. G. Harvey, R. A. Koch, C. D. McMahan, B. J. Nelson, C. E. Newman, and J. D. Satler. 2013. Model selection as a tool for phylogeographic inference: an example from the willow Salix melanopsis. Mol. Ecol. 22:4014-4028.

Choi, M., U. I. Scholl, W. Ji, T. Liu, I. R. Tikhonova, P. Zumbo, A. Nayir, A. Bakkaloğlu, S. Özen, and S. Sanjad. 2009. Genetic diagnosis by whole exome capture and massively parallel DNA sequencing. Proc. Nat. Acad. Sci. 106:19096-19101.

Chong, K. Y., C.-M. Chen, and K.-B. Choo. 1993. Post-hybridization recovery of membrane filter-bound DNA for enzymatic DNA amplification. Biotechniques 14:575-578.

Cock, P. J., T. Antao, J. T. Chang, B. A. Chapman, C. J. Cox, A. Dalke, I. Friedberg, T. Hamelryck, F. Kauff, and B. Wilczynski. 2009. Biopython: freely available Python tools for computational molecular biology and bioinformatics. Bioinformatics 25:1422-1423.

Crawford, N. G., B. C. Faircloth, J. E. McCormack, R. T. Brumfield, K. Winker, and T. C. Glenn. 2012. More than 1000 ultraconserved elements provide evidence that turtles are the sister group of archosaurs. Biol. Letters 8:783-786.

Davey, J. W., P. A. Hohenlohe, P. D. Etter, J. Q. Boone, J. M. Catchen, and M. L. Blaxter. 2011. Genome-wide genetic marker discovery and genotyping using next-generation sequencing. Nat. Rev. Genet. 12:499-510.

Delsuc, F., H. Brinkmann, and H. Philippe. 2005. Phylogenomics and the reconstruction of the tree of life. Nat. Rev. Genet. 6:361-375.

Eaton, D. A., and R. H. Ree. 2013. Inferring phylogeny and introgression using RADseq data: An example from flowering plants (Pedicularis: Orobanchaceae). Syst. Biol. 62:689-706.

Ellegren, H. In press. Genome sequencing and population genomics in non-model organisms. Trends Ecol. Evol.

Elshire, R. J., J. C. Glaubitz, Q. Sun, J. A. Poland, K. Kawamoto, E. S. Buckler, and S. E. Mitchell. 2011. A robust, simple genotyping-by-sequencing (GBS) approach for high diversity species. PLoS One 6:e19379.

Faircloth, B. C. 2011-12. Illumiprocessor. Available from https://github.com/faircloth-lab/illumiprocessor (last accessed December 20, 2013).

Faircloth, B. C., J. E. McCormack, N. G. Crawford, M. G. Harvey, R. T. Brumfield, and T. C. Glenn. 2012. Ultraconserved elements anchor thousands of genetic markers spanning multiple evolutionary timescales. Syst. Biol. 61:717-726.

Faircloth, B. C., L. Sorenson, F. Santini, and M. E. Alfaro. 2013. A phylogenomic perspective on the radiation of ray-finned fishes based upon targeted sequencing of ultraconserved elements (UCEs). PLoS One 8:e65923.

Felsenstein, J. 2006. Accuracy of coalescent likelihood estimates: do we need more sites, more sequences, or more loci? Molecular Biology and Evolution 23:691-700.

Fisher, S., A. Barry, J. Abreu, B. Minie, J. Nolan, T. M. Delorey, G. Young, T. J. Fennell, A. Allen, and L. Ambrogio. 2011. A scalable, fully automated process for construction of sequence-ready human exome targeted capture libraries. Genome Biol. 12:R1.

Fumagalli, M., F. G. Vieira, T. S. Korneliussen, T. Linderoth, E. Huerta-Sánchez, A. Albrechtsen, and R. Nielsen. 2013. Quantifying population genetic differentiation from next-generation sequencing data. Genetics 195:979-992.

Gladman, S. 2009. Velvet Optimiser. Available from http://bioinformatics.net.au/software.velvetoptimiser.shtml (last accessed December 20, 2013).

Glenn, T. C. 2011. Field guide to next-generation DNA sequencers. Mol. Ecol. Res. 11:759-769.

Gnirke, A., A. Melnikov, J. Maguire, P. Rogov, E. M. LeProust, W. Brockman, T. Fennell, G. Giannoukos, S. Fisher, C. Russ, S. Gabriel, D. B. Jaffe, E. S. Lander, and C. Nusbaum. 2009. Solution hybrid selection with ultra-long oligonucleotides for massively parallel targeted sequencing. Nat. Biotech. 27:182-189.

Gronau, I., M. J. Hubisz, B. Gulko, C. G. Danko, and A. Siepel. 2011. Bayesian inference of ancient human demography from individual genome sequences. Nat. Genet. 43:1031-1034.

Gutenkunst, R. N., R. D. Hernandez, S. H. Williamson, and C. D. Bustamante. 2009. Inferring the joint demographic history of multiple populations from multidimensional SNP frequency data. PLoS Genet. 5:e1000695.

Harris, J. K., J. W. Sahl, T. A. Castoe, B. D. Wagner, D. D. Pollock, and J. R. Spear. 2010. Comparison of normalization methods for construction of large, multiplex amplicon pools for next-generation sequencing. Appl. Environ. Microb. 76:3863-3868.

Harris, R. B., M. D. Carling, and I. J. Lovette. In press. The influence of sampling design on species tree inference: A new relationship for the New World chickadees (Aves: Poecile). Evolution doi:10.5061.

Hedtke, S. M., M. J. Morgan, D. C. Cannatella, and D. M. Hillis. 2013. Targeted enrichment: maximizing orthologous gene comparisons across deep evolutionary time. PLoS One 8:e67908.

Heled, J., and A. J. Drummond. 2010. Bayesian inference of species trees from multilocus data. Mol. Biol. Evol. 27:570-580.

Horner, D. S., G. Pavesi, T. Castrignanò, P. D. O. De Meo, S. Liuni, M. Sammeth, E. Picardi, and G. Pesole. 2010. Bioinformatics approaches for genomics and post genomics applications of next-generation sequencing. Brief. Bioinform. 11:181-197.

Huang, H., Q. He, L. S. Kubatko, and L. L. Knowles. 2010. Sources of error inherent in species-tree estimation: impact of mutational and coalescent effects on accuracy and implications for choosing among different methods. Syst. Biol. 59:573-583.

Hudson, R. R. 2002. Generating samples under a Wright-Fisher neutral model of genetic variation. Bioinformatics 18:337-338.

Jakobsson, M., S. W. Scholz, P. Scheet, J. R. Gibbs, J. M. VanLiere, H.-C. Fung, Z. A. Szpiech, J. H. Degnan, K. Wang, and R. Guerreiro. 2008. Genotype, haplotype and copy-number variation in worldwide human populations. Nature 451:998-1003.

Joshi, N. 2013. Sickle. Available from https://github.com/najoshi/sickle (last accessed December 20, 2013).

Kandpal, R. P., G. Kandpal, and S. M. Weissman. 1994. Construction of libraries enriched for sequence repeats and jumping clones, and hybridization selection for region-specific markers. Proc. Nat. Acad. Sci. 91:88-92.

Karagyozov, L., I. D. Kalcheva, and V. M. Chapman. 1993. Construction of random small-insert genomic libraries highly enriched for simple sequence repeats. Nucleic Acids Res. 21:3911.

Katoh, K., K. I. Kuma, H. Toh, and T. Miyata. 2005. MAFFT version 5: improvement in accuracy of multiple sequence alignment. Nucleic Acids Res. 33:511-518.

Katzman, S., A. D. Kern, G. Bejerano, G. Fewell, L. Fulton, R. K. Wilson, S. R. Salama, and D. Haussler. 2007. Human genome ultraconserved elements are ultraselected. Science 317:915-915.

Kuhner, M. K., J. Yamato, and J. Felsenstein. 1998. Maximum likelihood estimation of population growth rates based on the coalescent. Genetics 149:429-434.

Larget, B. R., S. K. Kotha, C. N. Dewey, and C. Ané. 2010. BUCKy: Gene tree/species tree reconciliation with Bayesian concordance analysis. Bioinformatics 26:2910-2911.

Lemmon, A. R., S. A. Emme, and E. M. Lemmon. 2012. Anchored hybrid enrichment for massively high-throughput phylogenomics. Syst. Biol. 61:727-744.

Li, C., M. Hofreiter, N. Straube, S. Corrigan, and G. J. Naylor. 2013. Capturing protein-coding genes across highly divergent species. Biotechniques 54:321-326.

Li, H. 2013. Seqtk. Available from https://github.com/lh3/seqtk (last accessed December 20, 2013).

Li, H., and R. Durbin. 2009. Fast and accurate short read alignment with Burrows-Wheeler transform. Bioinformatics 25:1754-1760.

Li, H., B. Handsaker, A. Wysoker, T. Fennell, J. Ruan, N. Homer, G. Marth, G. Abecasis, and R. Durbin. 2009. The Sequence Alignment/Map format and SAMtools. Bioinformatics 25:2078-2079.

Lynch, M. 2009. Estimation of allele frequencies from high-coverage genome-sequencing projects. Genetics 182:295-301.

Maddison, W. P. 1997. Gene trees in species trees. Syst. Biol. 46:523-536.

McCormack, J. E., B. C. Faircloth, N. G. Crawford, P. A. Gowaty, R. T. Brumfield, and T. C. Glenn. 2011. Ultraconserved elements are novel phylogenomic markers that resolve placental mammal phylogeny when combined with species-tree analysis. Genome Res. 22:746-754.

McCormack, J. E., M. G. Harvey, B. C. Faircloth, N. G. Crawford, T. C. Glenn, and R. T. Brumfield. 2013. A phylogeny of birds based on over 1,500 loci collected by target enrichment and high-throughput sequencing. PLoS One 8:e54848.

McCormack, J. E., J. M. Maley, S. M. Hird, E. P. Derryberry, G. R. Graves, and R. T. Brumfield. 2012. Next-generation sequencing reveals phylogeographic structure and a species tree for recent bird divergences. Mol. Phylo. Evol. 62:397-406.

Morin, R. D., M. Bainbridge, A. Fejes, M. Hirst, M. Krzywinski, T. J. Pugh, H. McDonald, R. Varhol, S. J. Jones, and M. A. Marra. 2008. Profiling the HeLa S3 transcriptome using randomly primed cDNA and massively parallel short-read sequencing. Biotechniques 45:81.

Narum, S. R., C. A. Buerkle, J. W. Davey, M. R. Miller, and P. A. Hohenlohe. 2013. Genotyping-by-sequencing in ecological and conservation genomics. Mol. Ecol. 22:2841-2847.

Nielsen, R., T. Korneliussen, A. Albrechtsen, Y. Li, and J. Wang. 2012. SNP calling, genotype calling, and sample allele frequency estimation from new-generation sequencing data. PLoS One 7:e37558.

Parchman, T., Z. Gompert, M. Braun, R. Brumfield, D. McDonald, J. Uy, E. Jarvis, B. Schlinger, and C. Buerkle. 2013. The genomic consequences of adaptive divergence and reproductive isolation between species of manakins. Mol. Ecol. 22:3304-3317.

Parker III, T. A., D. F. Stotz, and J. W. Fitzpatrick. 1996. Ecological and Distributional Databases Pages 113-436 *in* Neotropical Birds: Ecology and Conservation (D. F. Stotz, J. W. Fitzpatrick, T. A. Parker III, and D. Moskovits, eds.). University of Chicago Press, Chicago, Illinois.

Peterson, B. K., J. N. Weber, E. H. Kay, H. S. Fisher, and H. E. Hoekstra. 2012. Double digest RADseq: An inexpensive method for *de novo* genotyping in model and non-model species. PLoS One 7:e37135.

Pritchard, J. K., M. Stephens, and P. Donnelly. 2000. Inference of population structure using multilocus genotype data. Genetics 155:945-959.

Rambaut, A., and A. J. Drummond. 2007. Tracer v.1.5. Available from
http://tree.bio.ed.ac.uk/software/tracer (last accessed December 20, 2013).

Rambaut, A., and N. C. Grass. 1997. Seq-Gen: an application for the Monte Carlo simulation of
DNA sequence evolution along phylogenetic trees. Computer applications in the
biosciences: CABIOS 13:235-238.

Reid, N. M., S. M. Hird, J. M. Brown, T. A. Pelletier, J. D. McVay, J. D. Satler, and B. C.
Carstens. In press. Poor fit to the multispecies coalescent is widely detectable in
empirical data. Syst. Biol. doi:10.1093.

Ronquist, F., M. Teslenko, P. van der Mark, D. L. Ayres, A. Darling, S. Höhna, B. Larget, L.
Liu, M. A. Suchard, and J. P. Huelsenbeck. 2012. MrBayes 3.2: E, last accessed
December 20, 2013fficient Bayesian phylogenetic inference and model choice across a
large model space. Syst. Biol. 61:539-542.

Rubin, B. E., R. H. Ree, and C. S. Moreau. 2012. Inferring phylogenies from RAD sequence
data. PLoS One 7:e33394.

Salichos, L., and A. Rokas. 2013. Inferring ancient divergences requires genes with strong
phylogenetic signals. Nature 497:327-331.

Sboner, A., X. J. Mu, D. Greenbaum, R. K. Auerbach, and M. B. Gerstein. 2011. The real cost of
sequencing: higher than you think. Genome Biol 12:125.

Siepel, A., G. Bejerano, J. S. Pedersen, A. S. Hinrichs, M. Hou, K. Rosenbloom, H. Clawson, J.
Spieth, L. W. Hillier, and S. Richards. 2005. Evolutionarily conserved elements in
vertebrate, insect, worm, and yeast genomes. Genome Res. 15:1034-1050.

Smith, B. T., M. G. Harvey, B. C. Faircloth, T. C. Glenn, and R. T. Brumfield. 2014. Target capture and massively parallel sequencing of ultraconserved elements for comparative studies at shallow evolutionary time scales. Syst. Biol. 63:83-95.

Stolle, E., and R. F. A. Moritz. 2013. RESTseq: Efficient benchtop population genomics with restriction fragment sequencing. PLoS One 8:e63960.

Thornton, K. 2003. libsequence: a C++ class library for evolutionary genetic analysis. Bioinformatics 19:2325-2327.

van Rossum, G., and F. L. Drake (eds). 2001. Python Reference Manual. Python Labs, Virginia. Available from http://www.python.org (last accessed December 20, 2013).

Wagner, C. E., I. Keller, S. Wittwer, O. M. Selz, S. Mwaiko, L. Greuter, A. Sivasundar, and O. Seehausen. 2013. Genome-wide RAD sequence data provide unprecedented resolution of species boundaries and relationships in the Lake Victoria cichlid adaptive radiation. Mol. Ecol. 22:787-798.

Wang, S. E. Meyer, J. K. McKay, and M. V. Matz. 2012. 2b-RAD: A simple and flexible method for genome-wide genotyping. Nat. Methods 9:808-810.

Wang, Z., M. Gerstein, and M. Snyder. 2009. RNA-Seq: a revolutionary tool for transcriptomics. Nat. Rev. Genet. 10:57-63.

Weir, J. T., and D. Schluter. 2008. Calibrating the avian molecular clock. Mol. Ecol. 17:2321-2328.

Wetterstrand, M. S. 2013. DNA sequencing costs: Data from the NHGRI genome sequencing program (GSP). Available from http://www.genome.gov/sequencingcosts (last accessed December 20, 2013).

White, T. A., S. E. Perkins, G. Heckel, and J. B. Searle. 2013. Adaptive evolution during an ongoing range expansion: the invasive bank vole (*Myodes glareolus*) in Ireland. Mol. Ecol. 22:2971-2985.

Zerbino, D. R., and E. Birney. 2008. Velvet: Algorithms for de novo short read assembly using de Bruijn graphs. Genome Res. 18:821-829.

Table 1. Empirical Dataset Characteristics

| | RAD-Seq | | Sequence capture of ultraconserved elements | |
|---|---|---|---|---|
| | Sequences | SNPs | Sequences | SNPs |
| Number of loci | 4251 | 4251 | 1368 | 1262 |
| Mean locus length (bp) | 59.7 | 1 | 604.2 | 1 |
| Mean polymorphic sites per locus | 1 | 1 | 5.6 | 1 |
| % missing data | 0.2%[a] | 11.1% | 13.3% | 14.6% |
| Analyses conducted | Species tree | Population assignment | Species tree | Population assignment |
| | Demographic modeling (full coalescent) | Demographic modeling (allele frequency spectrum) | Demographic modeling (full coalescent) | Demographic modeling (allele frequency spectrum) |

[a] only reflects missing data in SNP site, remaining sequence is from tag consensus

Table 2. Population Genetic Summary Statistics from Empirical Data

| | SNPs (sample-wide)[a] | | | | Sequences (mean (SD))[b] | | | |
| | RAD-Seq | | Seq-cap | | RAD-Seq | | Seq-cap | |
| | trans-Andes | cis-Andes | trans-Andes | cis-Andes | trans-Andes | cis-Andes | trans-Andes | cis-Andes |
|---|---|---|---|---|---|---|---|---|
| Watterson's theta | 388.4 | 605.9 | 141.2 | 156.2 | 0.00313 (0.00365) | 0.00337 (0.00354) | 0.00155 (0.00195) | 0.00225 (0.00321) |
| Nucleotide diversity | 438.4 | 634.3 | 136.7 | 145.9 | 0.00345 (0.00423) | 0.00352 (0.00399) | 0.00157 (0.00215) | 0.00225 (0.00363) |
| Tajima's D | 0.708 | 0.257 | -0.173 | -0.361 | 0.33455 (0.89959) | 0.14555 (0.96385) | 0.02619 (0.86427) | -0.07452 (0.77634) |

[a] Values from SNPs are sample-wide, based on pairwise allele frequency spectrum for trans- and cis-Andean populations

[b] Values from sequences summarize variation across locus-specific estimates

Table 3. Demographic Parameter Estimates from Empirical Data

| | Ancestral $N_e$ | Trans-Andes $N_e$ | Cis-Andes $N_e$ | Trans->cis $N_{em}$ | Cis->trans $N_{em}$ | Substitution rate (/site/yr) |
|---|---|---|---|---|---|---|
| | Parameter ML estimates (bootstrap SD) from $\partial a \partial i$ | | | | | |
| RAD-Seq | 4,930,000 (1,888,311) | 221,000 (158,783) | 249,000 (169,987) | $1.36 \times 10^{-19}$ ($1.05 \times 10^{-9}$) | $2.27 \times 10^{-15}$ ($1.13 \times 10^{-10}$) | $4.56 \times 10^{-10}$ ($1.63 \times 10^{-10}$) |
| Seq-cap | 2,530,000 (1,851,795) | 64,900,000 (22,883,814) | 144,000,000 (45,921,473) | $2.61 \times 10^{-22}$ ($3.19 \times 10^{-6}$) | $1.33 \times 10^{-16}$ ($3.20 \times 10^{-6}$) | $1.74 \times 10^{-12}$ ($2.01 \times 10^{-11}$) |
| RAD-Seq | 9,420,000 (12,901,346) | 10,300,000 (29,762,194) | 9,640,000 (27,766,190) | $1.24 \times 10^{-39}$ ($4.69 \times 10^{-9}$) | $3.45 \times 10^{-22}$ ($7.57 \times 10^{-9}$) | $6.69 \times 10^{-12}$ ($1.17 \times 10^{-11}$) |
| Seq-cap | 2,180,000 (11,206,891) | 3,250,000 (14,646,088) | 65,400,000 (353,842,406) | $1.18 \times 10^{-9}$ ($1.67 \times 10^{-8}$) | $3.97 \times 10^{-11}$ ($2.81 \times 10^{-9}$) | $2.32 \times 10^{-11}$ ($5.17 \times 10^{-10}$) |
| | Parameter posterior mean (hpd interval) from G-PhoCS | | | | | |
| RAD-Seq | 3,158,671 (2,946,945 - 3,365,356) | 1,385,777 (1,286,526 - 1,486,926) | 1,591,807 (1,477,368 - 1,710,698) | $2.27 \times 10^{-4}$ ($1.22 \times 10^{-8}$ - $6.79 \times 10^{-4}$) | $2.09 \times 10^{-4}$ ($5.29 \times 10^{-9}$ - $6.30 \times 10^{-4}$) | $3.82 \times 10^{-10}$ ($3.52 \times 10^{-10}$ - $4.12 \times 10^{-1}$ |
| Seq-cap | 920,759 (840,964 - 1,002,497) | 1,229,287 (1,167,782 - 1,291,864) | 2,384,474 (2,253,559 - 2,415,518) | $9.58 \times 10^{-4}$ ($5.34 \times 10^{-8}$ - $2.32 \times 10^{-3}$) | $2.91 \times 10^{-4}$ ($4.59 \times 10^{-9}$ - $8.45 \times 10^{-4}$) | $3.73 \times 10^{-10}$ ($3.57 \times 10^{-10}$ - $3.90 \times 10^{-1}$ |

Table 4. RAD-Seq vs. Seq-cap for Systematics.

| | RAD-Seq[b] | Seq-cap |
|---|---|---|
| Investment | | |
| Equipment required | Thermal cycler, qPCR machine | Sonicator (or fragmentase), thermal cycler, rare earth magnet, qPCR machine |
| Supplies (library preparation) | $5-10 | $30-50 |
| Sequencing modal cost (typical values)[a] | ~$25 ($2-$25) | ~$25 ($25-$50) |
| Laboratory effort (library preparation) | ~ 2 days | 3-5 days |
| Computational requirements | Cluster or high-performance desktop | Cluster or high-performance desktop |
| Data characteristics | | |
| Efficiency | ~10% (taxon dependent) | ~30% (probe set dependent) |
| Number of loci | ≥Thousands | Hundreds to thousands |
| Mean locus length | Usually ≤200, up to 1000 possible | Usually 500-1000; >1000 possible |
| Repeatability | Low | High |
| Primary errors | Low coverage, paralogs | Low coverage |
| Primary applications | | |
| | Genome Scans | Phylogenetics |
| | Population genetics | Comparative phylogeography |
| | Single-species phylogeography | Studies of candidate loci |
| | Shallow phylogenetics | Studies using ancient DNA |
| | | Population Genetics |

Table S1. Sample Information

| | Museum | Tissue # | Biogeog. Area | Subspecies | Country | State | Locality | Lat. | Long. |
|---|---|---|---|---|---|---|---|---|---|
| 1 | KUMNH | 2044 | C. America | *X. m. mexicanus* | Mexico | Campeche | Calakmul, El Arroyo, 6 km S Silvituc | 18.5928 | -90.2561 |
| 2 | LSUMZ | 60935 | C. America | *X. m. mexicanus* | Honduras | Cortés | Cerro Azul Meamber National Park, Los Pinos | 14.8728 | -87.9050 |
| 3 | LSUMZ | 2209 | Chocó | *X. m. littoralis* | Panama | Darién | Cana on E slope Cerro Pirré | 7.7560 | -77.6840 |
| 4 | LSUMZ | 11948 | Chocó | *X. m. littoralis* | Ecuador | Esmeraldas | El Placer | 0.8667 | -78.5500 |
| 5 | LSUMZ | 4244 | Napo | *X. m. obsoletus* | Peru | Loreto | Lower Rio Napo, E bank Rio Yanayacu, ca. 90 km N Iquitos | -2.8200 | -73.2738 |
| 6 | LSUMZ | 6862 | Napo | *X. m. obsoletus* | Peru | Loreto | 5 km N Amazon River, 85 km NE Iquitos | -3.4167 | -72.5833 |
| 7 | LSUMZ | 9026 | Inambari | *X. m. obsoletus* | Bolivia | Pando | Nicolás Suarez, 12 km by road S Cobija, 8 km W on road to Mucden | -11.4703 | -68.7786 |
| 8 | FMNH | 433364 | Inambari | *X. m. obsoletus* | Peru | Cusco | Consuelo, 15.9 km SW Pilcopata | -13.0167 | -71.4833 |

NOTE: Museums are University of Kansas Natural History Museum (KUNHM), Field Museum (FMNH), and Louisiana State University Museum of Natural Science (LSUMZ).

Table S2. Comparison of $\partial$a$\partial$i Optimization Methods

| Optimization method | Mean Likelihood (20 replicates) |
|---|---|
| bfgs | -11288.15214 |
| l-bfgs-b | -9906.469736 |
| fmin | -31420.73871 |

Table S3. STRUCTURE Results With Different Values of *K*

| K | Pr(X\|K)[a] | Pr(K)[b] |
|---|---|---|
| RAD-Seq | | |
| 1 | -32473.5 | -31537.6 |
| 2 | -21191.8 | -19152.8 |
| 3 | -21268.8 | -19159.7 |
| 4 | -21342 | -19165.5 |
| 5 | -21391.7 | -19171.2 |
| 6 | -21454 | -19176.2 |
| 7 | -21504.8 | -19181.8 |
| 8 | -21550.5 | -19186.6 |
| Sequence capture | | |
| 1 | -8112.3 | -7826.6 |
| 2 | -6147.7 | -5479.8 |
| 3 | -5691.2 | -4837.4 |
| 4 | -12185.3 | -4015.2 |
| 5 | -46489.3 | -3515.1 |
| 6 | -16360.1 | -2852.1 |
| 7 | -121636.1 | -3047.4 |
| 8 | -51518.1 | -2929.4 |

[a]log probability of the data

[b]mean value of the log likelihood

Table S4. Mean Parameter Estimates Based on Simulated Datasets in ∂a∂i.

| Number of Loci | Ancestral Ne | Daughter 1 Ne | Daughter 2 Ne | Divergence Time (Yrs) | Migration Rate (1->2)[a] | Migration Rate (2->1)[a] |
|---|---|---|---|---|---|---|
| 500 | 168816 | 230161 | 262565 | 225092 | 8.12E-08 | 7.94E-08 |
| 1000 | 169496 | 296760 | 332266 | 232696 | 1.52E-07 | 1.38E-07 |
| 5000 | 312737 | 324176 | 321132 | 403355 | 2.22E-08 | 1.37E-08 |
| 10,000 | 277370 | 329419 | 317695 | 352773 | 2.03E-08 | 1.93E-08 |
| 50,000 | 344646 | 344145 | 358458 | 469799 | 4.26E-08 | 4.04E-08 |
| Expected (simulated) values | 390625 | 390625 | 390625 | 625000 | 6.40E-08 | 6.40E-08 |

[a] Number of individuals/year

Table S5. Detailed parameter estimates based on simulated datasets from G-PhoCS.

| | Ancestral Ne | Daughter 1 Ne | Daughter 2 Ne | Divergence Time (Yrs) | Migration Rate (1->2)[a] | Migration Rate (2->1)[a] |
|---|---|---|---|---|---|---|
| Number of Loci | | | | | | |
| 500 | 18931696 | 24357983 | 27391976 | 37755638 | 0.037121605 | 0.012987939 |
| 1000 | 19128001 | 27833450 | 30917407 | 35584246 | 0.042703123 | 0.012500745 |
| 5000 | 22212767 | 28127998 | 27779383 | 37768998 | 0.04240245 | 0.015727011 |
| 10,000 | 21434860 | 28895337 | 27823239 | 37452386 | 0.060610144 | 0.01420569 |
| 50,000 | | | | runs failed | | |
| Expected | 25000000 | 25000000 | 25000000 | 40000000 | 1.00E-09 | 1.00E-09 |
| Locus Length | | | | | | |
| 64 bp | | | see "500" treatment in Number of Loci above | | | |
| 500 bp | 62121875 | 57906250 | 65740625 | 77962500 | 0.093667243 | 0.026702156 |
| 1000 bp | 71631250 | 62946875 | 64431250 | 74312500 | 0.112703143 | 0.031040331 |
| 5,000 bp | 62862500 | 67440625 | 72515625 | 91400000 | 2.808480846 | 1.10949546 |
| 10,000 bp | 63690625 | 66662500 | 69206250 | 94112500 | 3.37660062 | 1.597286761 |
| Expected | 62500000 | 62500000 | 62500000 | 100000000 | 4.00E-10 | 4.00E-10 |

[a] Number of individuals/year

Figure 1. The expected distribution of reads produced by Illumina sequencing of DNA libraries produced by sequence capture and different RAD-Seq methods (inspired by Fig. 1 in Davey et al. 2011). Shaded areas depict reads, and the height of shaded regions indicates approximate expected read depths. Arrows point to restriction enzyme cut sites, and arrows that are black and white depict cut sites for different restriction enzymes. Information on sequence capture from , on original RAD-Seq from Davey et al. (2011), on GBS and RESTseq from Elshire et al. (2011) and Stolle and Moritz (2013), on ddRAD from Peterson et al. (2012), and on 2b-RAD from Wang et al. (2012). Actual results may vary.

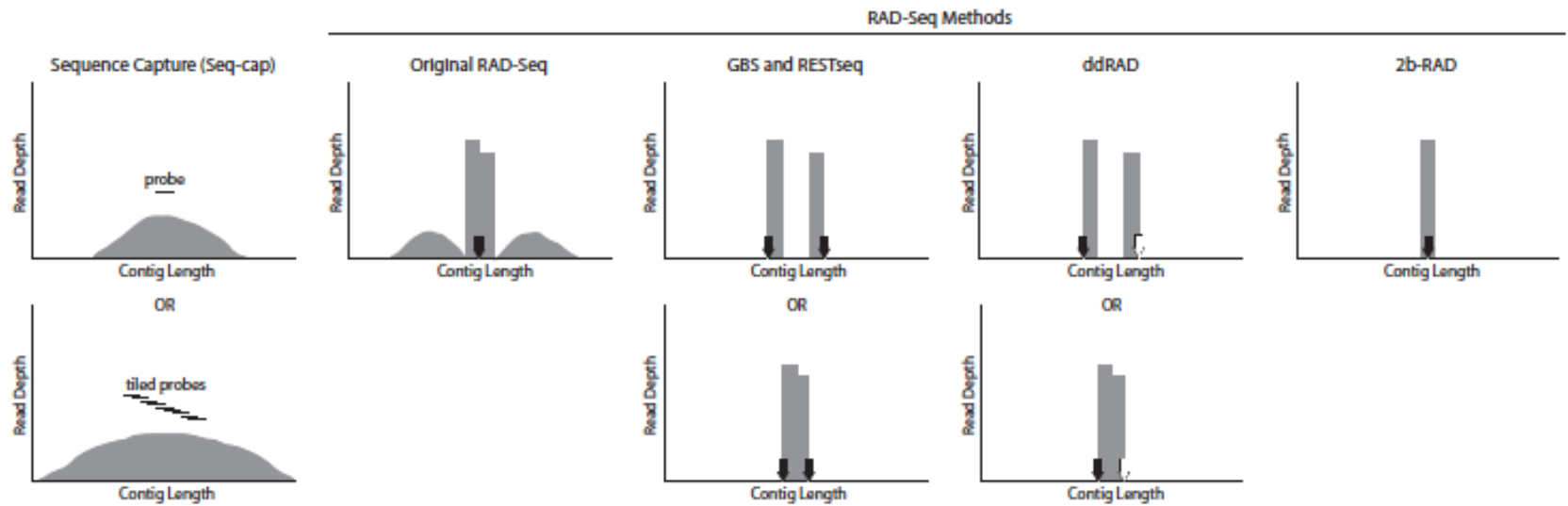

Figure 2. (a) Map showing the localities from which we sample individuals for this study, including four sites on either side of the Andes Mountains. (b) A depiction of the model used for demographic analyses with estimated parameters including thetas (θ), the population divergence time (τ), and migration rates (m). Trans-Andes refers to populations W of the Andes, whereas cis-Andes refers to populations E of the Andes.

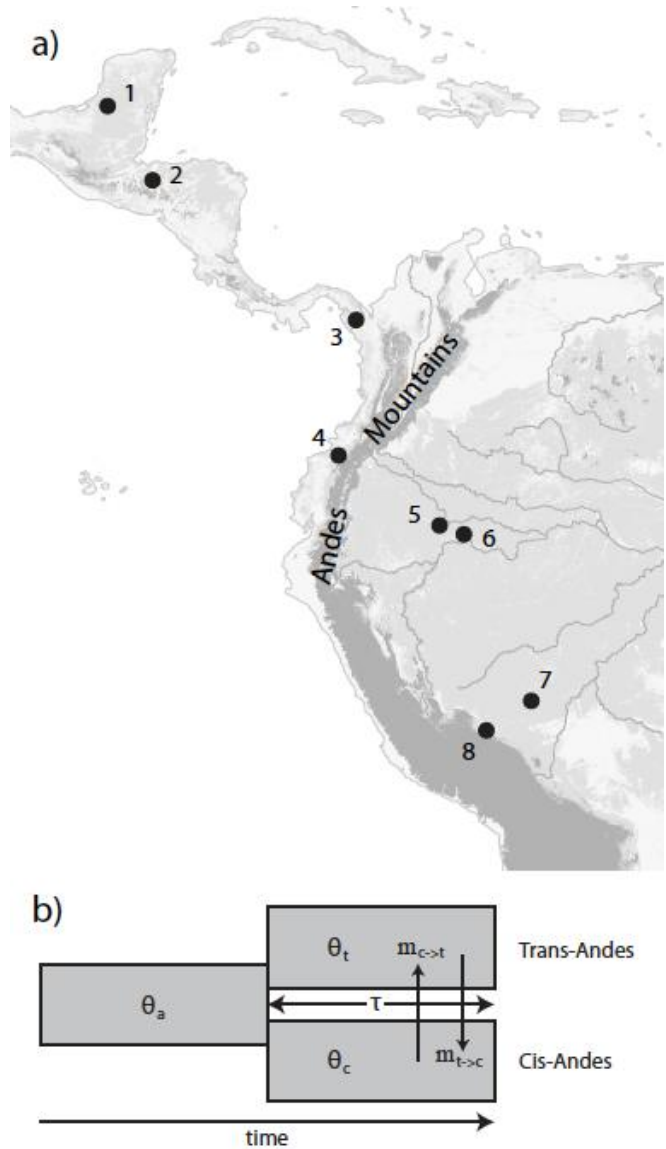

Figure 3. Phylogenetic trees showing the relationships between individuals based on RAD-Seq (a, b) and sequence capture (c, d) data. We show BUCKy primary concordance trees with concordance factors (a, c) and population trees with quartet concordance factors and branch lengths in coalescent units (b, d) for both datasets. We include a prior mitochondrial tree estimated in BEAST (e) for reference (Smith et al. in review).

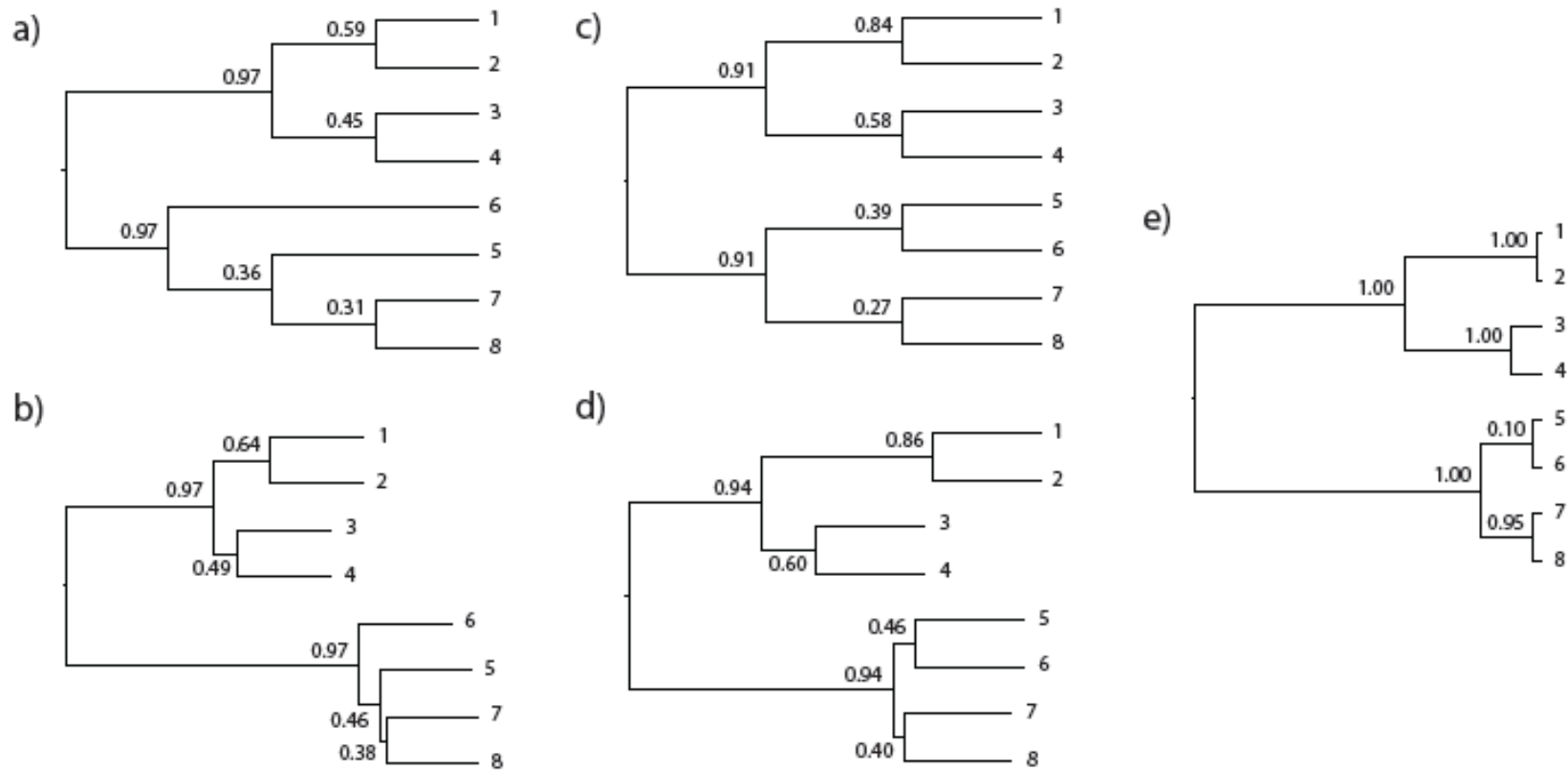

Figure 4. Simulation results including: (a) the influence of increasing locus length on quartet concordance factors for the split between two daughter populations in the BUCKy primary concordance tree, (b) the influence of increasing the number of loci on quartet concordance factors for the split between two daughter populations in the BUCKy primary concordance tree, (c) the influence of increasing the number of loci on the accuracy of parameter estimates (maximum likelihood value / expected value) in $\partial a \partial i$, and (d) the influence of increasing the number of loci on the precision of parameter estimates ((standard deviation/mean)$^2$) in $\partial a \partial i$. Migration rate estimates from $\partial a \partial i$ had low and variable accuracy and precision and are not depicted.

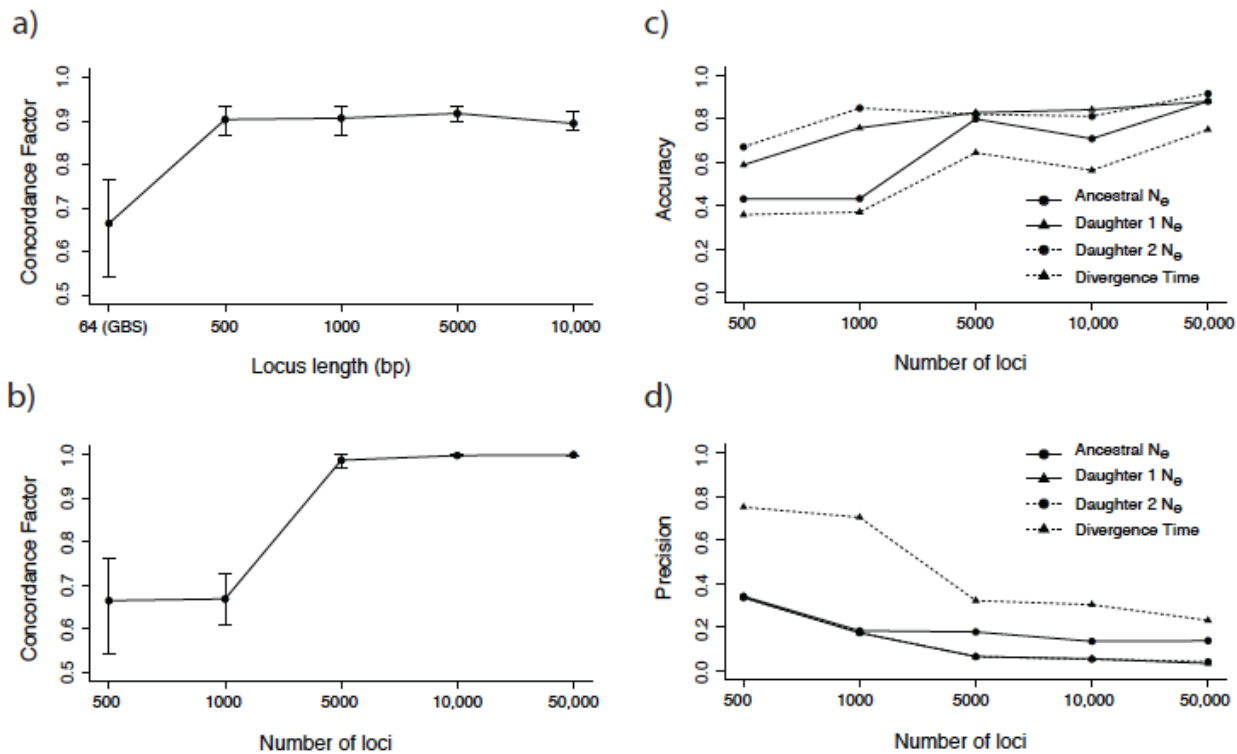